# An adaptive and fully automatic method for estimating the 3D position of bendable instruments using endoscopic images

Paolo Cabras · Florent Nageotte · Philippe Zanne · Christophe Doignon

**Department:**

ICube Laboratory, University of Strasbourg - CNRS
300, Bd. Sebastian Brant
CS 10413 - F-67412 Illkirch Cedex

**Responsible author:**
Paolo Cabras
1, Place de l'hôpital
67000 Strasbourg
Tel.: +33 3 88 11 91 29
p.cabras@gmail.com

**Financial Support:**

This work was supported by French state funds managed by the ANR within the Investissements d'Avenir program (Labex CAMI) under reference ANR-11-LABX- 0004.

**Category of manuscript:** original article
**Word count:**
**Number of figures:** 9

**Abstract.**

**Background.** Flexible bendable instruments are key tools for performing surgical endoscopy. Being able to measure the 3D position of such instruments can be useful for various tasks, such as controlling automatically robotized instruments and analyzing motions.

**Methods.** We propose an automatic method to infer the 3D pose of a single bending section instrument, using only the images provided by a monocular camera embedded at the tip of the endoscope. The proposed method relies on colored markers attached onto the bending section. The image of the instrument is segmented using a graph-based method and the corners of the markers are extracted by detecting the color transition along Bézier curves fitted on edge points. These features are accurately located and then used to estimate the 3D pose of the instrument using an adaptive model that allows to take into account the mechanical play between the instrument and its housing channel.

**Results.** The feature extraction method provides good localization of markers corners with images of *in vivo* environment despite sensor saturation due to strong lighting. The RMS error on the estimation of the tip position of the instrument for laboratory experiments was 2.1, 1.96, 3.18 mm in the x, y and z directions respectively. Qualitative analysis in the case of *in vivo* images shows the ability to correctly estimate the 3D position of the instrument tip during real motions.

**Conclusions.** The proposed method provides an automatic and accurate estimation of the 3D position of the tip of a bendable instrument in realistic conditions, where standard approaches fail.

**Keywords** Pose Estimation · *In vivo* image segmentation · Flexible endoscopy · Bendable instruments · Surgical Robotics

**1 Introduction**

For the last decade, flexible systems and bendable instruments have been used in new surgical interventions, such as intraluminal surgery. While their dexterity and adaptability to the human internal organs is suitable, their use for complex tasks is very demanding for surgeons. This has led to the development of complex manual platforms [1] and to robotic solutions. Robotic surgery has successfully addressed the limitations of minimally invasive surgery, by improving the easiness, time and accuracy of the procedures [2]. In our laboratory we have developed a teleoperation system for flexible endoscopic surgery, called STRAS, [3], which is based on the short Anubis flexible platform (Karl Storz, Tuttlingen, Germany) as presented in Fig. 1(a).

While this system and others, such as the IREP [4] developed at the Columbia University, have been successfully used in preclinical trial with teleoperation modes, the limited positioning accuracy of bendable instruments is still an important and recurrent issue [5], [6], [7]. It notably prevents the possibility to realize precise automatic positioning of the instruments. Even in teleoperated modes it requires the user to correct his/her inputs by constantly relying on the feedback from endoscopic images. It is therefore important to improve the positioning accuracy of flexible instruments for facilitating their use and to extend their capabilities in endoscopic surgery.

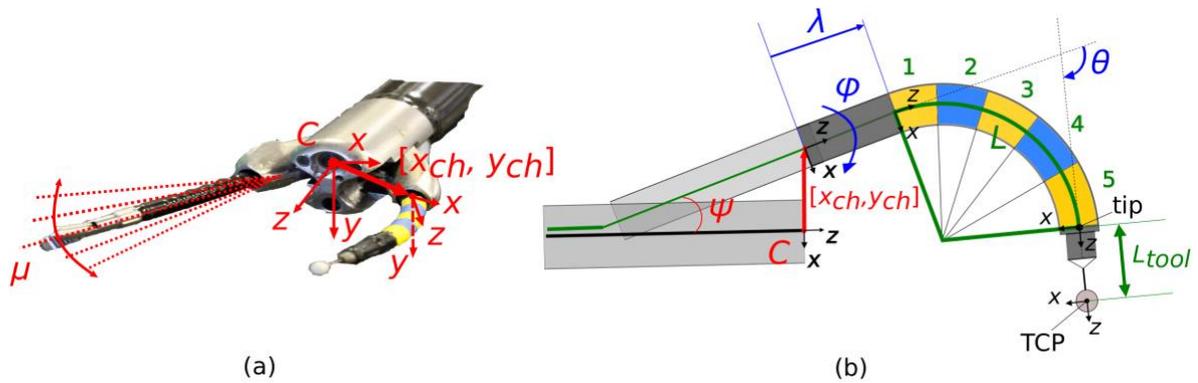

Fig. 1 In (a) a picture of the tip of the Anubis endoscopic platform with two bendable instruments emerging from the channels and in (b) the corresponding scheme with the parametric variables used for modeling the instruments.

The positioning accuracy of flexible bendable instruments is mainly hampered by the cable motion transmission from the actuators on the proximal side to the effector on the distal side. Cable transmission in flexible systems introduces highly non-linear effects such as backlash, hysteresis and dead-zones [8],[9]. Many mathematical and mechanical models have been proposed to represent these behaviors [10],[11]. However these models usually require the knowledge of many parameters, which are not readily available and difficult to identify. Moreover, because of hysteresis effects, the function that relates the position of the actuators to the position of the instruments can depend on a quite long history of motions performed by the instrument itself and the endoscopic guide, but also on external forces applied to them. The difficulty to comprehensively model the motion transmission in cable driven flexible systems suggests using sensors to measure the position of instruments. Additional sensors such as electromagnetic sensors [12], [13] or fiber Bragg grating have been suggested [14], but with difficulties due to the integration, sensitivity and *in vivo* compatibility of these sensors. Another approach could consist in using the video information provided by the endoscopic camera as proposed in [7]. However, these existing solutions require specific material such as stereoscopic cameras [5], [6] or have been applied to ad-hoc laboratory setup and cannot be directly used for real surgical systems [7]. Moreover these methods have never been applied *in vivo*. We have also explored this approach in [15]. However, features had to be selected manually, which made the approach unsuitable for processing image sequences on the fly. An approach based on learning techniques was also proposed in the same paper [15]. But such techniques require trustful ground truth for the training stage, which cannot be obtained easily.

In this paper, we present a complete scheme including image processing and computer vision methods, which automatically estimates the 3D position of the tip of a flexible single-section bendable instrument using a single embedded camera located at the tip of the endoscopic guide. Importantly, the methods are shown to work for *in vivo* images and for real working conditions of an endoscopic platform. The proposed approach, which relies on colored markers attached to the bending part of the instruments, is applicable to any single section bendable instruments, either with one or two planes of deflection, which can translate and rotate along their main axis. The scheme can be applied to manual or to robotized instruments with a large variety of uses, such as autonomous movements, gesture guidance or a posteriori gesture analysis.

This paper is structured as follows. Section 2 presents the models of the instruments and the image processing techniques proposed to robustly extract image features (apparent corners of the markers), as well as an original adaptive optimization process allowing to estimate the pose and configuration of the instruments. Section 3 reports results obtained on a laboratory setup and on *in vivo* images. The method and results are finally discussed in section 4.

## 2 Materials and Methods

Our objective is to propose a complete method for automatically estimating the position of the tip of single section bendable instruments, by relying mainly on the endoscopic images provided by the embedded endoscopic camera. This method should be compatible with real *in vivo* images. It should, therefore, include image processing steps for extracting useful 2D information from medical images and computer vision steps for inferring the 3D real position of the tip of the instrument or Tool Center Point (TCP) with respect to the endoscopic camera. By the term automatic we mean that the user does not have to interact with the software for tuning parameters or defining regions of interest in the endoscopic images.

### 2.1 Endoscopic system

As shown in Fig. 1(a), the distal part of the short Anubis platform (Karl Storz), that we use for experiments, is a surgical endoscopic system composed of an endoscopic camera and two instruments (diameter 3.5mm) coming out from two channels included in the body of the endoscope. These instruments have three degrees of freedom (DOFs): translation and rotation with respect to the channel axis and deflection (parameterized respectively by $\lambda$, $\varphi$ and $\theta$, see Fig. 1(b)). The bending part of the instrument is 18.5 mm long and consists of 12 vertebrae. Deflection is obtained by actuating a pair of antagonist cables running inside the shaft of the instrument and attached to the more distal vertebra. Such bending systems are usually modeled by an arc of a torus whose major radius can be controlled [9]. Note that the same geometrical modeling can be applied to bendable systems with two orthogonal bending planes [13].

For describing more accurately the system, four other parameters, referred as mechanical parameters, are used in our work: the position of the extremity of the instrument channel in the camera plane ($[x_{ch}, y_{ch}]$) expressed with respect to the origin of the camera and two Euler angles $\phi$ and $\mu$ representing the orientation of the channel axis with respect to the camera frame, around $y$ axis and $x$ axes respectively (see Fig. 1(a) and (b)). These four parameters are usually considered as fixed and known [5], but we will show in the following the practical need to consider them variable in order to model the motion of the instrument inside the channel, due to play. These four mechanical parameters are grouped in $r_m = [x_{ch}, y_{ch}, \psi, \mu]^T$. The complete configuration of one instrument with respect to the endoscopic camera is denoted $r = [x_{ch}, y_{ch}, \psi, \mu, \lambda, \varphi, \theta]$.

The geometric structure of the bendable instruments is assumed known (length of bendable section, distance from the end of the bendable section to the effector). We also assume that the intrinsic parameters of the endoscopic camera have been obtained by a calibration process [16]. It must be noted that strong radial distortion effects have to be taken into account on flexible endoscopic systems.

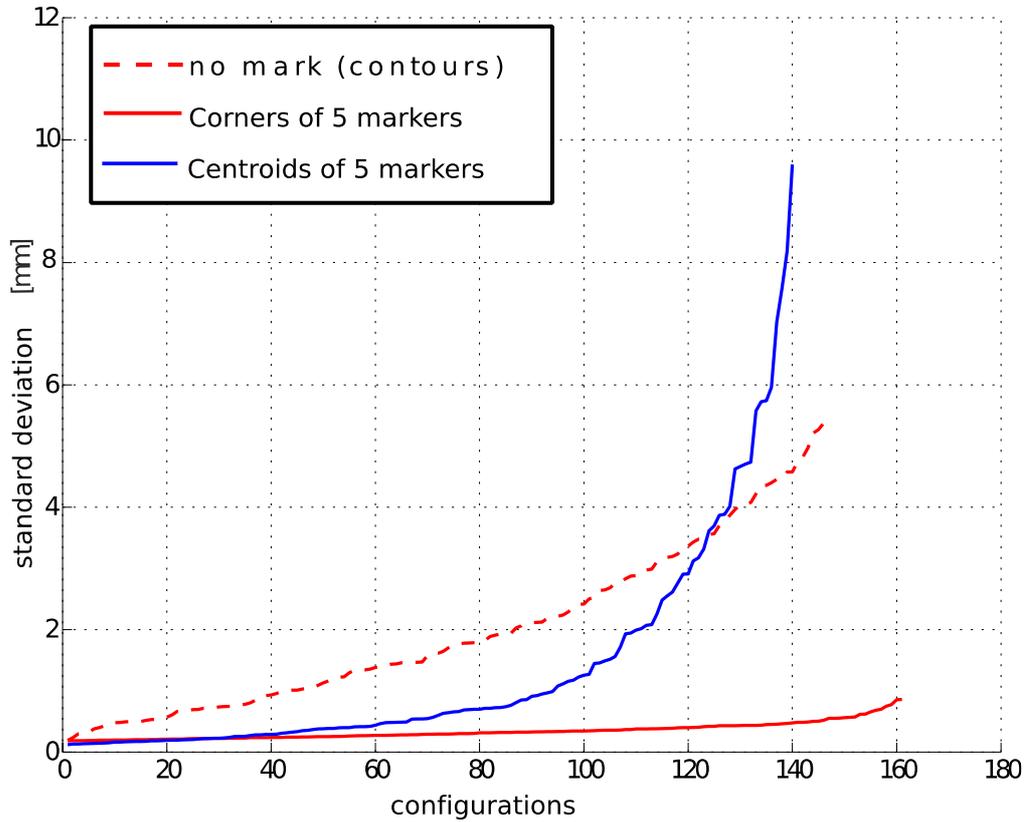

Fig.2 Simulated estimation errors (in mm) on the tip of the instrument position for 160 configurations of one instrument obtained for different features: no markers (use of apparent contours only) (dashed red), apparent corners of five markers (red), centroids of 5 markers (blue). The errors are obtained by propagating a one pixel standard deviation noise onto the features up to the 3D position of the instrument. The errors are then sorted for comparison purposes.

## 2.2 Choice of visual features and markers

The objective of this work is to measure the 3D position of the TCP with respect to the endoscopic camera. One assumption for our approach is that the tip of the bendable instruments (i.e. the effector) is often in interaction with tissues, and can therefore be hidden. Consequently, our choice is to rely on the image of the bendable section of the instruments, which is in general at least partly visible, and to estimate the position of the TCP through the 3D configuration of the bendable section.

For estimating the pose of a deformable object a single image, approaches such as shape from shading or shape from template could be considered. However, we have chosen to rely on the use of well-chosen discrete features, in order to provide robustness to occlusions and to be independent on lighting conditions monitoring

Two aspects have then to be considered: to select adequate features for capturing the shape of the object and computing the pose of the bendable section, and to extract them from the image of the instrument itself or from artificial markers attached to it. Both aspects are linked, since increasing the number of features on a defined space usually increases theoretical pose estimation accuracy, but at the cost of making extraction more difficult. Given the generally bad quality of *in vivo* images provided by conventional flexible endoscopes (limited resolution, smoke and fluids in the field of view) and the need of simple and robust features extraction, we have chosen to attach color markers to the bendable part of the instrument. To assure the visibility of markers whatever the rotation of the instrument and for providing robustness to partial occlusion (which may be caused by fluids or because of important specular effects as it will be shown in section 3.3), we have opted for several monochrome stripes rolled around the bendable section of the instrument, providing a ruler-like pattern (Fig. 1(a)). Rolled markers have been used in laparoscopic surgery before [17] and this kind of ruler-like pattern is also used on conventional gastroenterology instruments such as Needle Knives (Olympus, Tokyo, Japan) to provide users with good motion perception. For assuring good visibility and distinguishability between them, the number of markers has been limited to five for the considered instruments (length =18.5 mm, diameter =3.62mm). Markers colors have been chosen in order to be able to discriminate them from the mostly red and pink *in vivo* background and from each other. We have thus opted for an alternation of blue and yellow markers of equal sizes (3.7 mm long).

To our knowledge, there is no demonstrated optimal choice of features for estimating the pose of bendable instruments. Therefore, we have carried out a theoretical analysis in order to choose suitable features. For this purpose the sensitivity of the estimation of the 3D position of the instrument TCP with respect to features extraction errors has been analyzed for different types of features. Given the previous choice of markers, three configurations have been considered: (1) instrument apparent contours (or silhouette, which could also be used without markers), (2) markers centroids and (3) markers apparent corners. Contours and marker corners are said "apparent" because they do not have constant physical counterparts on the real instrument. The physical counterparts slide onto the surface of the object when the point of view is changed, as for the limb points of spheres seen from a perspective projection.

The results of this analysis are summarized in Fig. 2, which shows the ordered errors on the TCP for a large set of configurations of the instrument distributed onto its workspace. This study shows that apparent contours and centroids are inadequate because of their high sensitivity to noise. Much higher robustness can be expected with apparent corners of markers.

## 2.3 Features Extraction

The extraction of the apparent corners of the markers is quite challenging. The conventional approaches (e.g. Harris detector, maximum curvature of contours, etc.) are not efficient because the object has no sharp vertexes and the images are of low quality. The apparent corners can be defined as the points corresponding to the color transitions on the upper (or lower) instrument apparent contour. To exploit this property, we propose to first roughly segment the instrument in the image using colors so as to define a region of interest where the contours will be subsequently detected. In order to reject falsely segmented areas, an interpretation step is necessary (see section 2.3.1), which sorts the markers from the base to the tip. A continuous definition of the upper and lower ordered border points is then obtained so as to be able to detect the color transitions with a subpixel accuracy. This process also allows to label the extracted corners, which makes the matching with the 3D landmarks possible (see Fig. 3 for the overview of the whole process).

We now describe in details the interpretation step (section 2.3.1) and the border fitting method (section 2.3.2), which are the key parts of this image processing, before giving a complete overview of the scheme (section 2.3.3).

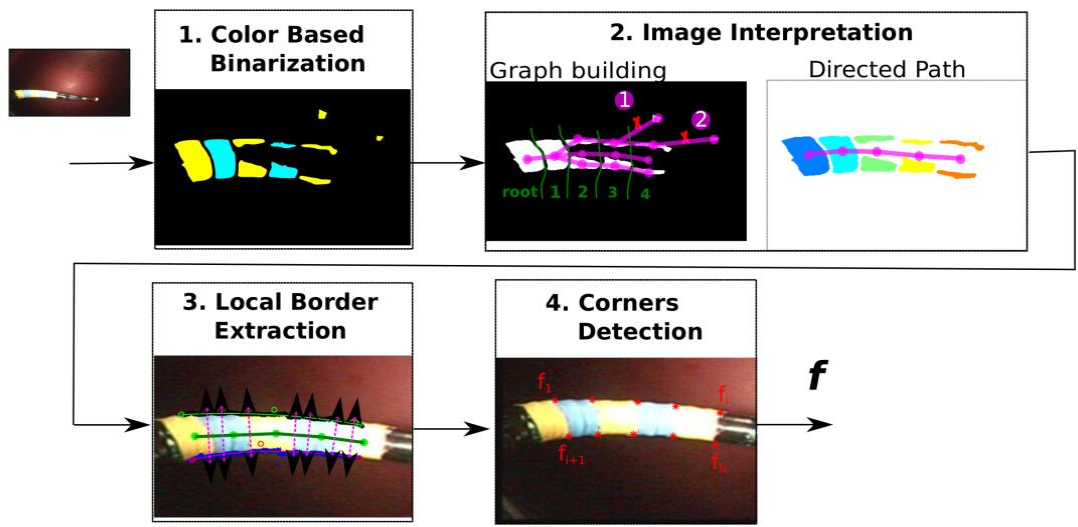

Fig. 3: Steps of the features extraction process. (1) GM are used to select and label blue and yellow regions. (2) Creation of the tree using the specified connectivity: Branch 2 is not created since it does not fulfill condition 1, whereas branch 1 does not respect condition 2 because the orientation of this branch strongly differs from the preceding outline. The tree is then processed to get a directed path. (3) The skeleton is used for searching border points. These are subsequently fitted with Bézier curves. (4) Color discontinuities are searched along the Bézier curves and considered as corners.

**2.3.1 Graph-Based Image Interpretation**

A first coarse segmentation can be realized based on colors. However, because of the strong illumination coming from the endoscope, large specular areas often appear onto the instrument, that remove color information and split markers projections into several regions (see step 1 of Fig. 3). The candidate regions have also to be labeled according to the marker number they belong to (from 1 to 5) or classified as false detections (which can arise because of reflections onto the organs for instance). This problem is solved in two steps by using a graph-based method: firstly, a tree is created according to the regions connectivity and, secondly, it is processed to obtain a directed path, where the nodes represent each marker from the base (the root of the tree) to the tip.

2.3.1.1 Tree construction

The connectivity between regions used for creating the tree is defined based on two structural characteristics:

1. The markers are adjacent to one another alternating between blue and yellow,

2. The skeleton linking the centroids of the subsequent markers should not exhibit large changes of directions. The limits have been empirically chosen to ±60º.

After building the tree using this connectivity criteria, each level $i$ of the tree contains all the candidate regions associated to marker $i$ ($i$=1 for the first marker, $i$=5 for the last marker). For each level, the nodes are also considered by pairs and virtual nodes, representing the regions resulting from the merging of both nodes are created. This provides an extended tree as shown in Fig. 4.

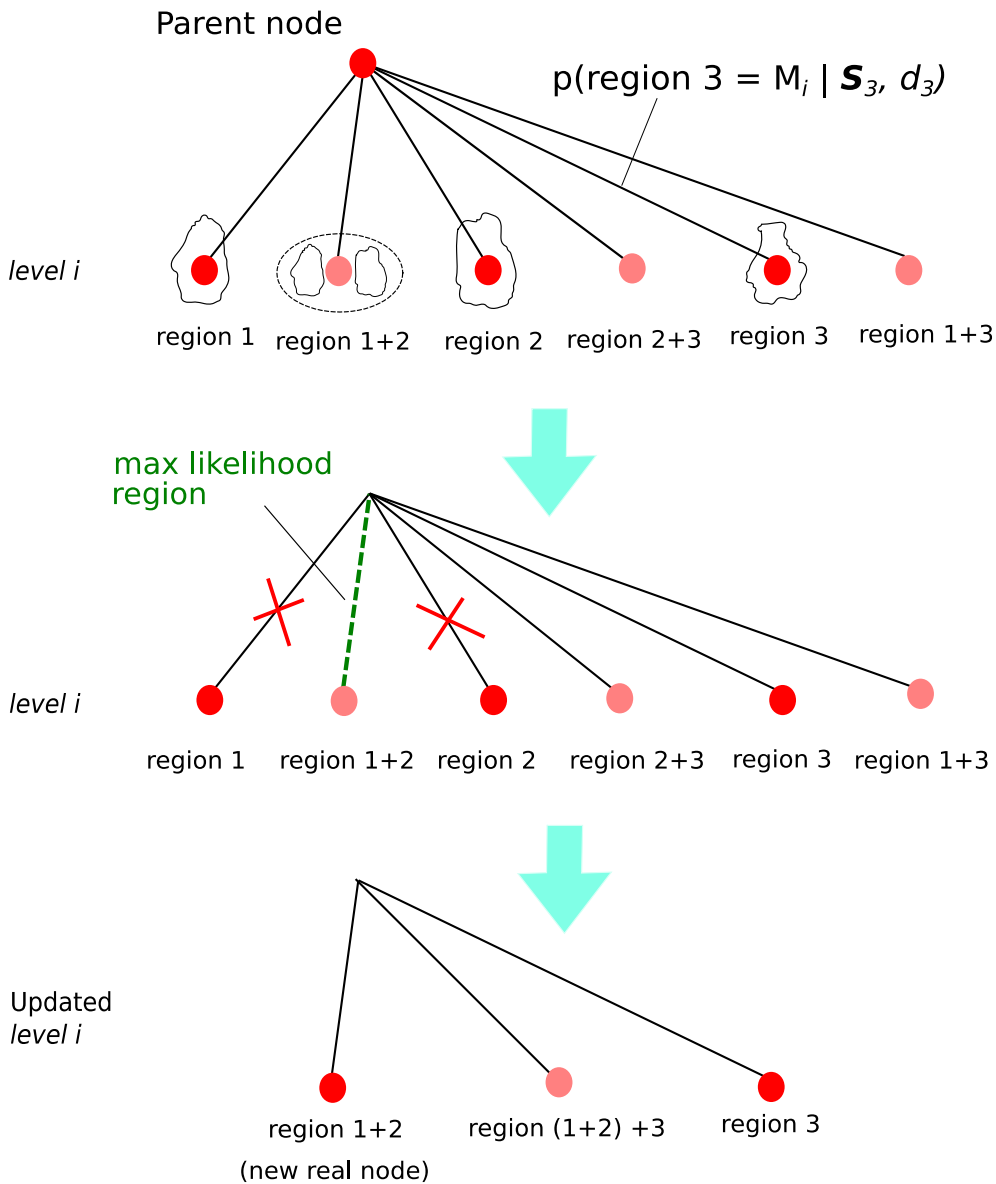

Fig. 4: Scheme of the tree updating process. At each level, the branch with the highest value (maximum likelihood) is selected (highlighted in green). If, as in the presented case, the branch

reaches a virtual node (i.e. the node representing the merge of two regions), the nodes of the two original regions are deleted. If the level contains more than two regions, new merging actions are considered, branches values are re-computed and the process is repeated on the updated tree (third line).

2.3.1.2 From tree to path

Then the correct directed path has to be extracted from the tree. For this purpose, we rely on the appearance of the markers and their relative distances. For each node of the tree, one defines two characteristics (Fig. 5):

- a 2 components vector $\mathbf{S}$, which contains the minor and major axis half lengths of the region associated to the node,

- a scalar $d$, which is the distance of the considered node (real or virtual) from its parent node.

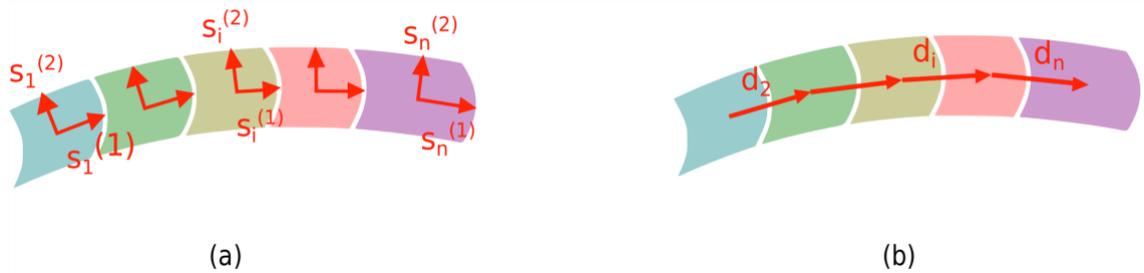

Fig. 5: The characteristics associated to each marker used for selecting the correct path in the tree: one shape factor and a topological factor. The former is defined as the length and width of each marker along its principal axes (left) and the latter as the distance from the considered marker to its parent (right).

A priori information on the appearance of the markers can then be used to remove incorrect nodes and select the correct ones. General assumptions (such as maximal distances or asymmetry of the candidates shape) are useful, but frame-specific a priori information may be needed to obtain a single path.

In the case of an image sequence, likelihoods on the shape ($p(S|M_i)$) and distance ($p(d|M_i)$) given a marker $M_i$ can be obtained, for a specific frame, by considering time-consistency. For instance, the previous values of the characteristics could be used or Kalman filtering could be designed. In the case of a robotized instrument, one could use the information from the proximal encoders to generate a likelihood model for the characteristics.

Then, for one candidate node $k$ at level $i$ with shape $S_k$ and distance $d_k$ to its parent node, the a posteriori probabilities of being marker $i$ (given the *shape* and *distance* evidences ) can be computed as follows (according to the Bayes theorem):

$$p(M_i|S_k, d_k) = p(M_i|S_k)\, p(M_i|d_k) \propto p(S_k|M_i)\, p(d_k|M_i) \qquad (1)$$

The a posteriori likelihoods are then assigned as values to the incoming branches of a node. The tree is then processed from root to leaf, level by level (see Fig. 4), by selecting the branch with the highest value for each level transition. If the chosen branch points to a virtual node, the tree structure is updated: the considered real nodes are merged and the weight values are recomputed for the branches affected by the merging action.

This approach also allows handling empty levels, but for the sake of clarity of the global process this case is not described here.

Likelihoods were also used in [7] for selecting correct candidates for markers on bendable instruments during tracking tasks. Characteristics that were used were distances from extracted regions to the past position of the markers and the ratio between the regions areas and the past markers areas. Since the distances are directly computed with respect to the markers positions in a previous image, matching can fail in the case of fast movements or a good prediction is needed. In our approach, the distance that is used is computed between candidate regions for successive markers in the current image and it is compared with the distance between markers in a previous image (image sequence) or with a prediction (robotized instruments). Since these inter-markers distances vary more slowly than the positions of markers during the movements of the instrument, our approach is arguably more robust to fast motions. Moreover and importantly, no mechanism for reconstituting markers from several regions were proposed in [7], which can make this approach sensitive to specularities encountered in *in vivo* images (see section 3).

### 2.3.2 Bézier Curve Representation of the Borders

At the end of step 3 of the image processing (see Fig. 3 and section 2.3.3), two ordered lists of detected border points are obtained using local maximum gradient extraction: One list for the upper contour of the instrument and one list for the lower contour. Parametric curve fitting can then be used to smooth contours, in order to provide sub-pixel precision, and help rejecting outliers. Tests performed on synthetic data have shown that a quadratic Bézier curve is sufficient to represent the apparent borders of the projection of a torus, from the particular point of view of the considered endoscopic system. Here we describe this parametric curve fitting for any list of ordered candidate border points (upper or lower), denoted by $\boldsymbol{p}_i = (p_i^{(x)}, p_i^{(y)})$.

A quadratic Bézier curve is defined as follows:

$$\mathbf{B}(t) = \sum_{k=0}^{2} b_k \boldsymbol{\beta}_k = \boldsymbol{\beta}_0 + t(-2\boldsymbol{\beta}_0 + 2\boldsymbol{\beta}_1) + t^2(\boldsymbol{\beta}_0 - 2\boldsymbol{\beta}_1 + \boldsymbol{\beta}_2) \qquad (1)$$

where

$$b_k = \binom{2}{k} t^k (1-t)^{(n-k)}, k = \{0,1,2\}. \qquad (2)$$

are the Bernstein basis polynomials of degree 2 and $\beta_k$ are the control points of the curve. The fitting problem can be defined as minimizing the following objective function:

$$Q = \sum_{i=0}^{n} \rho\left(\mathbf{B}_x(t_i, \boldsymbol{\beta}_x) - p_i^{(x)}\right) + \sum_{i=0}^{n} \rho\left(\mathbf{B}_y(t_i, \boldsymbol{\beta}_y) - p_i^{(y)}\right) \qquad (3)$$

on $\mathbf{a} = \left[\beta_{0x}, \beta_{1x}, \beta_{2x}, \beta_{0y}, \beta_{1y}, \beta_{2y}\right]^T$ (the $x$ and $y$ coordinates of the control points), $t_i$ is the value of the parameter of the Bézier curve and $\rho$ is a loss function used to ensure some robustness of the fitting with respect to outliers.

The coordinates of the points on the Bézier curve (2) at parameters values $t_i$ can be expressed in a matrix form:

$$\begin{pmatrix} \mathbf{B}_x \\ \mathbf{B}_y \end{pmatrix} = \mathbf{TCa} \qquad (4)$$

where $\mathbf{C}$ is the matrix of the Bézier coefficients,

$$\mathbf{C} = \begin{bmatrix} 1 & -2 & 1 & 0 & 0 & 0 \\ -2 & 2 & 0 & 0 & 0 & 0 \\ 1 & 0 & 0 & 0 & 0 & 0 \\ 0 & 0 & 0 & 1 & -2 & 1 \\ 0 & 0 & 0 & -2 & 2 & 0 \\ 0 & 0 & 0 & 1 & 0 & 0 \end{bmatrix}$$

and $\mathbf{T}$ is composed of lines $[t_i^2, t_i, 1, 0, 0, 0]$ for $x$ abscissas and $[0, 0, 0, t_i^2, t_i, 1,]$ for $y$ ordinates. The proposed cost function considers a point to point error, where the residuals consist of the differences of each coordinates (abscissas and ordinates) of corresponding points $\boldsymbol{q}_i(\boldsymbol{a}) = \left[ B_x(t_i, \mathbf{a}) - p_i^{(x)}, B_y(t_i, \mathbf{a}) - p_i^{(y)} \right]^T$. This requires attributing a value $t_i$ of the parameter to a considered image point $\mathbf{p}_i$. The control polygon of the Bézier curve, defined by points $\boldsymbol{\beta}_0, \boldsymbol{\beta}_1$ and $\boldsymbol{\beta}_2$ should begin and end at the terminal parts of the apparent (superior or inferior) contour of the bendable part of the instrument. Therefore, since the candidate contour points are ordered (see section 2.3.3), we compute $t_i$ as the curvilinear coordinate over the curve connecting the ordered points:

$$t_i = \frac{d_i}{d_n}$$

where:

$$\begin{cases} d_1 &= 0 \\ d_i &= d_{i-1} + \|\boldsymbol{p}_i - \boldsymbol{p}_{i-1}\|_2 \end{cases}$$

In case of non-quadratic loss functions, the problem becomes non linear but it can be solved by an Iterative Re-weighted Least Square method. At each iteration $(j + 1)$, then, the searched $\mathbf{a}^{(j+1)}$ is computed as

$$\mathbf{a}^{(j+1)} = \left( (\mathbf{TC})^T \mathbf{W}^{(j)} (\mathbf{TC}) \right)^{-1} (\mathbf{TC})^T \mathbf{W}^{(j)} \mathbf{q} \tag{5}$$

The weights on the diagonal of $\mathbf{W}^{(j)}$ can be updated according to the chosen loss function $\rho$. Thanks to this robust fitting technique, a continuous representation of the apparent contours can be obtained, which allows to filter out the outliers due to image noise or miss-detection.

### 2.3.3 Workflow of the whole process

To summarize, the algorithm employed to extract the apparent corners of the markers consists of one preliminary off-line stage and 4 on-line stages (see Fig. 3).

*Preliminary stage.* This stage is used for obtaining a coarse model of the color appearance of the markers in typical endoscopic images. This is performed off-line, typically only once (as long as markers colors are not modified). A few samples of the markers (Regions of Interest (ROI)) are manually selected from *in vivo* frames containing the image of the marked instrument. The colors in the ROI are expressed in the L*a*b color space and are used to construct two 2-D Gaussian models (one model for yellow and one model for blue) in the a*b color plane. This choice has been made because yellow and blue occupy opposite locations along the b axis.

*First stage: Candidate regions extraction.* Thanks to the off-line Gaussian models, the image to be processed is labeled with three categories (yellow / blue / background) according to two thresholds (one for each color).

*Second stage: Image Interpretation.* The marker associated to the base of the bendable section (the root of the tree) is generally well detected because it lies in a part of the workspace that does not receive much direct illumination. The tree is then built and processed till obtaining the sequence of the 5 centroids coordinates ordered from the base to the tip of the bendable section as explained in section 2.3.1.

*Third stage: Apparent Borders Detection.* The skeleton obtained by connecting the ordered centroids is evenly sampled. From each sample point, normals to the skeleton are drawn and local maxima of gradient magnitude are searched along these normals to the skeleton. The candidate contours points are selected according to their gradient direction (aligned with the normal to the skeleton) and their position (near the boundaries of the labeled image obtained in the previous step). The candidate with maximum gradient magnitude is kept for each normal and on each side of the skeleton. Bézier curves are then fitted onto the candidate contours as described in section 2.3.2. For robust fitting, we use the Beaton and Tukey loss function [18]. According to that, the updated weights for each coordinate of $\boldsymbol{q_i}$, are computed as:

$$w_i^{(j+1)}(q_i) = \begin{cases} \left[1 - \left(\frac{q_i/\sigma}{c}\right)\right], & if \ |q_i/\sigma| \leq c \\ 0, & if \ |q_i/\sigma| > c \end{cases}$$

where $q_i$ indicates either $q_i^{(x)}$ or $q_i^{(y)}$ and where the scale $\sigma$ is taken as the median absolute deviation (MAD) of the residuals. As suggested in [19], we update the scale value during the first 4 iterations and fix it afterwards (no appreciable variations occur after the 4 first iterations). Optimal c must be chosen according to the application. Here the value c = 1.5 has been empirically selected and appeared to be appropriate.

*Fourth stage: Corner Localization.* Since blue and yellow are complementary colors, the limit between two markers is defined as the intersection between the outlines of blue and yellow along the apparent borders (see Fig. 6b).

After this process, N ($\leq 12$) apparent corners are extracted, which will be used as features for the subsequent pose estimation process (see section 2.4). Four corners could be sufficient, but all the available extracted features are of course conserved to bring redundancy and robustness to features mislocation.

In the following $F_i$ will denote any 3D apparent corner (feature) whose image has been extracted. $F = [F_1, ..., F_N]^T$ (dimensions $3 \times N$ ) is the concatenation of the 3D coordinates of the apparent corners. The same notations but with a lowercase **f** will be used to express the coordinates of the same points projected in the image. The hat symbol (^) is used for estimated values.

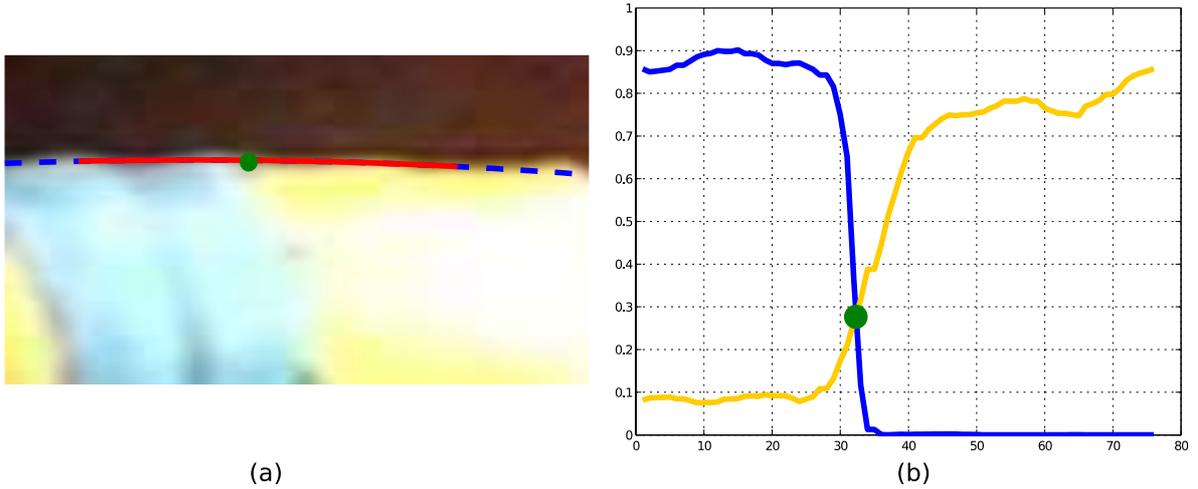

(a)                  (b)

Fig. 6: Along each Bézier curve representing the boundary of the instrument (left), the blue and yellow chroma profiles are computed (right) and the apparent corners are defined as the zero crossing points of the signal obtained by the point-to-point subtraction of the yellow and blue signals.

## 2.4 Model-Based 3D Pose Estimation

To the best of our knowledge, there is no closed-form solution for estimating the pose of bendable instruments from a single perspective projection. However, the problem of the 3D pose estimation can be formalized as an optimization procedure whose aim is to find the best instrument configuration in terms of a "cost function" based on image measurements [20].

As explained in our early work [15], considering that the mechanical parameters are known and fixed during the manipulation does not work for real surgical systems. Indeed, these parameters slightly vary during the surgical operation, mainly because of the mechanical play between the instrument and the channel. The mechanical play model presented in that work has been completed here by also considering variations on angle µ. Therefore the configuration of the instrument will be described by the extended vector of DOFs ($r$), which contains both the mechanical parameters and the conventional DOFs of the bendable instrument: $r = [x_{ch}, y_{ch}, \psi, \mu, \lambda, \varphi, \theta]$. Furthermore, the Gauss-Newton optimization method proposed in [15] may present convergence problems because of the high number of parameters used, which can be partly dependent in some configurations. This is why a Levenberg-Marquardt (L-M) approach is now used for the optimization, which allows to handle parameters redundancy.

### 2.4.1 Optimization Process

The basic optimization problem consists in finding the value $r_{opt}$ of the extended DOFs $r$, which minimizes the distance between the extracted image features $f$ and the virtual projection of the corresponding apparent corners in the configuration and pose parameterized with **r**, denoted $\hat{f}$. However, because of the high number of parameters, several minima where the value of the cost function is close to the global minimum may be found but with incorrect physical signification. It is therefore important to constrain the search space. For this purpose, it is reasonable to assume that the actual values of the mechanical parameters remain close to their nominal values $r_m^* = [x_{ch}^*, y_{ch}^*, \psi^*, \mu^*]^T$, which can be obtained from the Computer-Aided Design (CAD) model of the endoscopic system.

This a priori information is taken into account in the cost function (following the concept initially proposed by Lowe in [21]), by penalizing large deviations of each mechanical parameter from its nominal value. This is achieved with the following penalty function:

$$\rho(u) = \frac{k}{3}\left|\frac{u}{a}\right|^3, a \in \mathbb{R}^+, k \in \mathbb{R}^+ \qquad (6)$$

where $u$ can be any component of vector $\mathbf{u} = \mathbf{r}_m - \mathbf{r}_m^*$, $k$ determines the steepness of the function outside a "free zone" defined by half-width $a$. Different values of $k$ and $a$ can be used for different mechanical parameters.

The corresponding weight function used in the minimization process is given by

$$\zeta(u) = \frac{k}{a^3}|u|, a \in \mathbb{R}^+, k \in \mathbb{R}^+. \qquad (7)$$

The cost function, then, becomes:

$$\chi^2 = \frac{1}{2}\|\boldsymbol{f} - \hat{\boldsymbol{f}}\|^2 + \rho_{ch}(x_{ch}^* - \hat{x}_{ch}) + \rho_{ch}(y_{ch}^* - \hat{y}_{ch}) + \rho_\psi(\psi^* - \hat{\psi}) + \\ +\rho_\mu(\mu^* - \hat{\mu}). \qquad (8)$$

This problem can be solved iteratively according to a L-M approach. Defining:

- $\mathbf{J}_{g_i}$ the geometric Jacobian so that $\dot{\mathbf{F}}_i = \mathbf{J}_{g_i}\dot{\mathbf{r}}$
- $\mathbf{J}_{I_i}$ and $\mathbf{L}_i$ so that $\dot{\mathbf{f}}_i = \mathbf{J}_{I_i}\mathbf{J}_{g_i}\dot{\mathbf{r}} = \mathbf{L}_i\dot{\mathbf{r}}$ (similarly to the visual-motor jacobian in [22])
- $\mathbf{L}$ the matrix obtained by stacking $\mathbf{L}_i$, and given an estimate, the correction to be added to the current value of $\mathbf{r}$ to decrease $\chi^2$ is given by:

$$\delta\mathbf{r} = (\mathbf{L}^T\mathbf{W}^T\mathbf{W}\mathbf{L} + \varepsilon\mathbf{I})^{-1}\mathbf{L}\mathbf{W}^T\mathbf{W}\mathbf{e} \qquad (9)$$

where the error vector is given by

$$\mathbf{e} = \left[(\mathbf{r}_m^* - \hat{\mathbf{r}}_m)^T, \quad (\mathbf{f} - \hat{\mathbf{f}})^T\right]^T.$$

$\varepsilon$ is the damping factor of the L-M approach determined on the basis of the cost evolution. $\mathbf{W}$ is a weighting matrix allowing to take into account the penalty on the mechanical parameters values:

$$\mathbf{W} = \begin{bmatrix} \zeta(e_1) & 0 & 0 & 0 & \\ 0 & \zeta(e_2) & 0 & 0 & \\ 0 & 0 & \zeta(e_3) & 0 & \mathbf{0}_{4\times N} \\ 0 & 0 & 0 & \zeta(e_4) & \\ & & \mathbf{0}_{N\times 4} & & \mathbf{I}_{N\times N} \end{bmatrix}.$$

The expressions of Jacobian and interaction matrices are omitted here but they can be found in [23].

This formalism provides the optimization with the ability to automatically adapt to changes on the device structure and to be robust to mechanical play. Note that the same formalism could be used with other features.

To prove the effectiveness of the entire pose estimation algorithm, it was tested on both a manual device and a robotic device as presented hereafter.

## 3 RESULTS

### 3.1 Experimental setups

The proposed method was tested on a laboratory setup in order to obtain quantitative information and *in vivo* where no ground truth was available.

The laboratory setup consists of the STRAS robotic platform and a stereoscopic system looking at the instrument workspace. All cameras have been calibrated and registered using planar calibrations grids [16]. The acquisitions of all images are synchronized. The endoscopic images have PAL format. The background is reddish in order to mimic real conditions. The accuracy of the estimation is evaluated as the 3D distance between the position of the effector of the instrument (TCP) estimated from image features, and the 3D Ground Truth (GT).

The GT is provided by the stereo system and expressed in the endoscopic camera frame. The maximal uncertainty of the GT has been assessed to 0.91 mm, with RMS uncertainty of 0.49 mm, 0.30 mm and 0.28 mm respectively for *x*, *y* and *z* camera axes. The TCP (white ball in Fig. 1(a)) is distant 15.8 mm from the end of the bendable part (this is almost as long as the bendable part itself). Using this measurement point for assessment allows to take into account the quality of the estimation of both the position and orientation of the bending part.

*In vivo* images have been obtained with a manual version of the Anubis platform. This system is therefore slightly different from a geometrical point of view.

For both cases, the nominal geometric parameters given by the CAD model for the left channel are $r_m^* = [x_{ch}^*, y_{ch}^*, \psi^*, \mu^*]^T = [-13.3 \text{ mm}, 6.2 \text{ mm}, 10°, 0°]$. On the basis of numerous simulations considering model uncertainties, the parameters of the penalty functions $a, k$ (c.f. eq. (7)) have been chosen as $k_\psi = k_\mu = 2, a_\psi = a_\mu = 1°$ and $k_{ch} = 15, a_{ch} = 1$ mm (cf. eq. (8) and (9)).

The Gaussian models for the color appearance of blue and yellow markers (preliminary stage of section 2.3.3) have been obtained by manually selecting the markers in 20 frames randomly selected from *in vivo* sequences. These models are computed only once and they are used for all experiments, including laboratory experiments.

For all experiments, the a priori used for extracting the directed path in the tree representation of candidate regions (see section 2.3.1) is obtained in the following manner. For each marker $M_i$, each topological factor (shape and distance) is considered as a Gaussian variable and the mean and standard deviation of these variables are estimated using the 5 last available images where the instrument was extracted. This is a basic approach, which allows to take into account the variation of the apparent shape of the instrument when the processed frames are obtained from a temporal sequence.

### 3.2 Laboratory experiment

The laboratory experiment was carried out on the robotic version of the Anubis platform (STRAS). The complete proposed method has been tested on a set of 295 images. The configurations were chosen so as to cover the useful portion of the instrument workspace that is visible from the camera and they were obtained by evenly sampling the range of movement of each motor actuating the instrument 3 degrees of freedom. The root mean square (RMS) errors over each coordinate (in mm) are 2.1 ± 1.05 on x (horizontal axis of camera), 1.96 ± 1.71 on y (vertical axis of camera) and 3.18 ± 2.14 on z axis (depth with respect to camera).

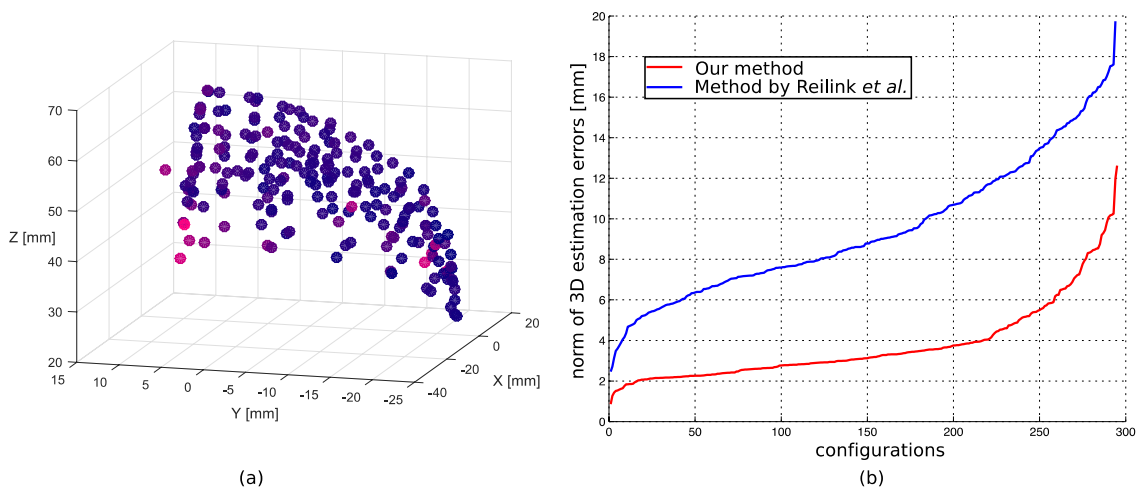

Fig. 7: In (a), representation of the 3D TCP positions of the validation sequence and the associated error: the more reddish the higher is the norm of the estimation 3D error. In (b), sorted norms of the 3D errors on the TCP estimation using variable (our method, red) or fixed (method of [5], blue) mechanical parameters.

The obtained results are better than the results of our early work reported in [15] for the same robotic endoscopic system and similar feature points (note that the points were selected manually in [15]). This improvement first confirms the effectiveness of the segmentation method and may also point out the interest of letting $\mu$ vary and the adoption of a Levenberg-Marquardt optimization approach. These results can also be compared with the approach proposed in [7] where the fixed nominal model is used during the pose estimation. If we use this method for the same extracted features, errors become more than two times larger (in mm): 3.92 ± 2.47, 3.61 ± 1.85 and 9.37 ± 4.91 for the same 3 axes. In these conditions, our method outperforms the method proposed in [7]: almost half of the errors are beneath 3mm for our method vs 9 mm for [7] and 85% below 5 mm vs only 5% for [7]. This supports the usefulness of the adaptability of the pose estimation scheme for taking into account mechanical play.

It can be noted that the errors obtained by applying the method of [5] onto the extracted features, provide larger position errors than what is directly reported in [5]. This discrepancy probably mainly accounts for the fact that we apply the method on a realistic endoscopic setup (similar endoscopic systems have been used for complete manual and robotic procedures onto animal models), where significant play between instruments and channels exists in order to allow low friction translation and rotation of the instruments, whereas the system used in [5], while similar in its form, is an ad-hoc laboratory endoscopic system. In addition to this aspect, using a validation point (the white ball) far from the features used for the pose estimation amplifies the errors committed on the orientation and reflects them on the position error. In [7], the validation point was the tip of the instrument next to the corners used for the pose estimation, hence partly masking the effect of potential errors on the orientation.

By examining the spatial distribution of the errors (see Fig. 7(a)), it can be seen that the largest errors are located in two areas of the workspace corresponding to a deflected instrument directed downward and close to the camera. This may be caused by the fact that, in these configurations, the instrument is close to the border of the image, where distortion effects due to large field of view lenses are the most important. This intrinsically increases the sensitivity of the estimation process. Moreover, camera distortion estimation is also probably less accurate in this area.

### 3.3 *In vivo* results

The same process has been applied on two *in vivo* video sequences, acquired in the abdomen of a porcine model during single port laparoscopic surgery realized with the manual short Anubis platform. The first sequence, 3 minutes long, was acquired near the intestine, while the second sequence, also 3 minutes long, was acquired near the liver region, thus providing different backgrounds. No ground truth for the 3D TCP position was available for these *in vivo* images.

In Fig.8 and 9 some typical images are presented for both sequences. The top line shows the detection of the instruments apparent contours (red points) and the extracted apparent corners (green crosses) obtained as described in section 2.3. The 3D pose and configuration of the instrument $\mathbf{r_{opt}}$ is then computed based on the extracted corners as explained in section 2.4. A virtual image of the instrument is then computed by projecting the instrument in the estimated pose $\mathbf{r_{opt}}$ onto the image (equivalent to $\hat{f}$). Red crosses on the bottom line of Fig. 8 and 9 show the position of the virtual instrument contours and the green dot shows the estimated position of the center of the grasper (TCP).

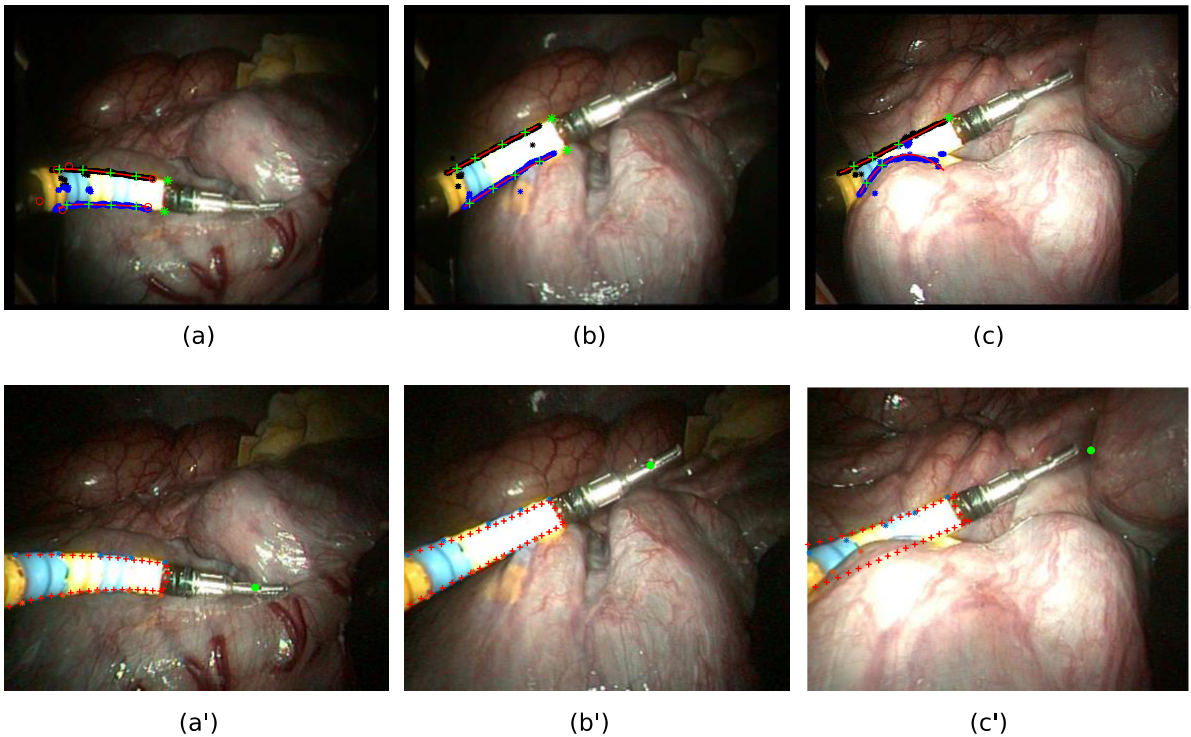

Fig. 8: *In vivo* results of the presented method with intestine in the background: (Top images) the features extraction process and (bottom images) the virtual projection of the instrument from the estimated pose (red crosses). The green dot is the projection of the center of the grasper. In (a-a'), the real instrument is correctly discriminated from its reflection on the organ. In (b-b'), the adaptive capability of the algorithm allows to fit the instrument even when it is pushing on the organ. In (c-c') the segmentation process partly fails for the lower border because the instrument is hidden.

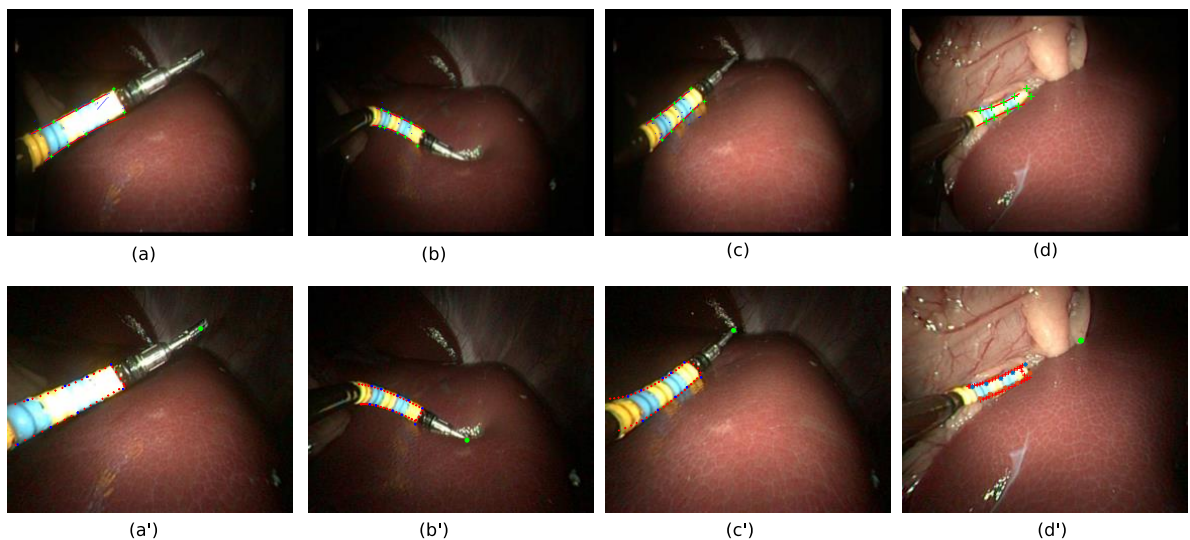

Fig. 9: *In vivo* results with liver in the background: (Top images) the segmentation process and (bottom images) the virtual projection of the instrument from the estimated pose (red crosses). The green dot is the projection of the center of the grasper. In (a-a'), the real instrument is correctly segmented despite strong saturation, in (c-c') it is correctly discriminated from its reflection on the organ. The adaptive capability of the algorithm allows to fit the instrument

even when it is pushing on the organ (b-b'). In (d-d') an example of the result obtained when the effector is hidden.

A qualitative analysis can then be carried out by comparing the real image with the virtual image. It shows that the complete proposed method is capable of segmenting the instrument and the markers corners even in presence of strong light saturation and mirroring effect on the surface of the organ (Fig. 8 (b), Fig. 9 (a) and (c)). The consideration of mechanical plays in the model through the variation of the mechanical parameters allows the pose estimation process to converge even when the instrument is pushing on the organs, which modifies its position inside the channel and deforms the flexible shaft of the instrument (Fig. 8 (a) and (a'), Fig. 9 (b) and (b')). However, in this case, the estimation of the position of the tip is slightly offset because of the deformation of the bendable section itself.

A complete qualitative analysis of the features extraction process has also been carried out for the whole sequences. For each frame, features extraction is considered successful if at least 80% of the visible corners are correctly detected. One then gets success rates of 86.7% for the intestine region and 72.7% for the liver region. The algorithm fails when the instrument is partially or totally hidden: in this case the apparent borders are not visible and cannot be correctly detected, leading to an obvious misdetection of the corners. Particular environmental conditions (mainly due to light interaction with the environment or presence of liquid on the camera lens) can also affect the perception of the markers, hence modifying their apparent shape and leading to a mislocation of the color transition.

## 4 DISCUSSION

We have presented a fully automatic technique to retrieve the 3D position of the TCP of marked bendable robotized instruments using a single endoscopic image and a coarse a priori on the configuration of the instrument. The method includes image features extraction compatible with *in vivo* environment and an adaptive pose estimation process for the bendable section of the instrument, which allows to handle unavoidable change of position of the instruments in their channels.

The features extraction method focuses on feature saliency and spatial coherency and it is based on a graph analysis. It is able to reject the numerous outliers that can appear in *in vivo* environments and can reliably extract and match instrument apparent corners even if data is partly missing. Importantly, the graph-based interpretation of the image allows to reconstruct markers made of several extracted regions and allows to cope with specularities. To our knowledge, this is the first marker extraction method for bendable instruments, which is successfully used on *in vivo* images.

Markers images labeling and matching with the 3D landmarks is an important stage. Indeed, fitting the instrument model using only the apparent contours information would lead to a cost function with many minima corresponding to the configurations where the projected (virtual) apparent contour is included in the actual apparent contour of the instrument but does not cover it completely. This is also the reason why, as shown in section 2.2 and Fig. 2, apparent contours are not sufficient as visual features for pose estimation. Indeed, to properly fit the model, points with 3D significance are needed and, to extract such points, they have to be labeled according to their position with respect to the instrument. Moreover, in an *in vivo* environment, due to the possible presence of several moving organs and the peculiar strong illumination, the instrument contours extraction can be difficult and should be limited to a particular zone. *In vivo* experiments have shown that even with partial information only, the computed 3D pose and configuration are acceptable in terms of reprojection errors.

The features extraction process works on individual images. Only a coarse estimation of the instrument configuration is needed for complex cases (for instance images with instrument reflection onto the organs) to select the correct path in the graph. In the case of a robotic system this estimation can be obtained from forward position kinematic models fed with encoder data. In the case of manual systems, time consistency can be used. In our experiments, the same a priori has been used for laboratory experiments and for *in vivo* experiments, whereas the motions were very different: In the laboratory experiments the workspace was discretized, hence creating large motions between successive frames. For the *in vivo* experiments both rapid and slow motions were performed with the instruments. This demonstrates that only rough a priori information is necessary and that simple models are sufficient. Obviously, other techniques such as Kalman filtering could also be used.

Laboratory experiments were performed to compare the instrument tip 3D position computed with the embedded endoscopic image with that provided by an external stereovision system observing simultaneously the instrument. These tests showed that admitting tolerances over the mechanical parameters allows to achieve quantitative results better than the state of the art methods in estimating the position of the TCP of the instrument. Better numerical results were reported in our previous work [15] for similar laboratory conditions. But in that case a learning-based approach was used, which inherently presents several disadvantages with respect to the presented technique. Firstly, it requires a trustful ground truth (hard and time-consuming to obtain). Moreover, the learning-based approach does not provide the ability to adapt itself to even slight changes in the system, meaning that a specific training set should be obtained for each particular system. For instance different training sets would be needed for the STRAS robot and the manual Anubis platform since their mechanical parameters are practically different.

On the contrary, one key feature of the proposed model-based approach is its ability to adapt itself and work for different conditions without any tuning of parameters as shown in the experiments. Indeed, among these experiments, two different endoscopic devices were used, different configurations of the endoscopic guide were encountered and different backgrounds were considered.

The proposed method allows to obtain the position of the tip of the instrument considering a single image. Temporal filtering, the exploitation of multi-frames information and tracking could be added in the future to improve the accuracy of the estimation in time sequences. Nevertheless, the current method will still be necessary for recovering after instrument disappearance, occlusion or tracking failure.

Quantitative *in vivo* validation would be beneficial, but using stereo images would require a very complex setup. EM sensors are not trustful for absolute measurements given compatibility issues with the operating room. Nevertheless, quantitative differences between *in vivo* and laboratory data treatment will only arise from the features extraction stage. Therefore, since, from a qualitative point of view, features are extracted as precisely for *in vivo* images as for laboratory images, one can reasonably assume that similar position accuracies could be obtained.

Images analysis shows that errors can still be large in some cases despite good features extraction. Improvements could maybe be obtained by relaxing some hypotheses such as the constant curvature of the bendable section.

The surgical instruments used in the experiments have a single bending plane. However the same method can be applied to instruments with two orthogonal bending planes, such as the ones used in [13]. Indeed, the position of the TCP can be parameterized the same manner and the appearance of the projection in the image is identical. Therefore the features extraction process and the estimation of the 3D position of the instrument TCP remain valid for such instruments.

## 5 CONCLUSION

We have presented a complete method for estimating the position of bendable instruments using a single endoscopic camera embedded at the tip of the flexible endoscopic system. The approach relies on two important contributions: (1) an algorithm for segmenting marked bendable instruments in endoscopic images, whose workflow is detailed in section 2.3.3. It has been tested on long images sequences acquired *in vivo*, and it has been shown to work without any parameter tuning for two different environments and despite difficult conditions, providing success rates of 72% and 86%. To our knowledge this is the first method reported, which has been successfully tested onto *in vivo* images; and (2) an adaptive pose estimation process, which takes into account imperfections of real endoscopic systems. Laboratory experiments were performed to provide quantitative assessment for the whole method and better results were obtained than those provided by the existing state-of-the art methods.

The whole technique or any of its two main parts can be used for all single section bendable instruments (two or four ways bending) used with standard endoscopes or in surgical platforms, either in manual applications or in motorized setups. The proposed framework can be used in many applications, such as automatic instrument positioning, assistance to teleoperation or a posteriori motion and gesture analysis.

Concerning extensions of this work, we are interested in analyzing the ability of the segmentation process to deal with the interaction of two instruments. The robustness and maybe the accuracy of the proposed technique could potentially be improved by using temporal tracking in the case of image sequences or by combining proximal measurements with image estimation in the case of robotic endoscopes. Concerning the pose estimation process, more complex instrument models, for instance with non-uniform curvature, could also be considered. The impact of such models on the convergence and the generalization to different instruments should then be analyzed. Concerning applications, our current work is currently focused on the use of the position estimation in a visual servoing scheme for controlling robotic flexible instruments.

# Conflicts of interest notification page

This statement is to certify that not one of the authors have conflict of interest to declare. The authors certify that they have NO affiliations with or involvement in any organization or entity with any financial interest or non-financial interest in the subject matter or materials discussed in this manuscript.

**Funding**. This work was supported by French state funds managed by the ANR within the Investissements d'Avenir program (Labex CAMI) under reference ANR-11-LABX-0004.